\documentclass[%
reprint,
superscriptaddress,
 amsmath,amssymb,
 aps,
 prb,
]{revtex4-2}

\usepackage[colorlinks, linkcolor=blue, anchorcolor=blue, citecolor=blue, urlcolor=blue]{hyperref} 
\usepackage{graphicx}
\usepackage{float}

\begin{document}

\title{Gate tuning of coupled electronic and structural phase transition in atomically thin Ta$_2$NiSe$_5$}

\author{Keyu~Wei}
\affiliation{National Laboratory of Solid State Microstructures and Department of Physics, Nanjing University, Nanjing 210093, China}

\author{Yixuan Luo}
\affiliation{State Key Laboratory of Quantum Functional Materials, School of Physical Science and Technology, ShanghaiTech University, Shanghai 201210, China}

\author{Kenji Watanabe}
\affiliation{Research Center for Electronic and Optical Materials, National Institute for Materials Science, Tsukuba 305-0044, Japan}

\author{Takashi~Taniguchi}
\affiliation{Research Center for Materials Nanoarchitectonics, National Institute for Materials Science, Tsukuba 305-0044, Japan}

\author{Yanfeng Guo}
\email{guoyf@shanghaitech.edu.cn}
\affiliation{State Key Laboratory of Quantum Functional Materials, School of Physical Science and Technology, ShanghaiTech University, Shanghai 201210, China}
\affiliation{ShanghaiTech Laboratory for Topological Physics, ShanghaiTech University, Shanghai 201210, China}

\author{Xiaoxiang~Xi}
\email{xxi@nju.edu.cn}
\affiliation{National Laboratory of Solid State Microstructures and Department of Physics, Nanjing University, Nanjing 210093, China}
\affiliation{Collaborative Innovation Center of Advanced Microstructures, Nanjing University, Nanjing 210093, China}
\affiliation{Jiangsu Physical Science Research Center, Nanjing 210093, China}

\begin{abstract}
Realizing an excitonic insulator phase from narrow-gap semiconductors remains challenging, as unambiguous experimental signatures are difficult to establish. Ta$_2$NiSe$_5$ has been widely regarded as a leading candidate, yet the nature of its phase transition and insulating state remains controversial. Here, we report a systematic Raman spectroscopy study of Ta$_2$NiSe$_5$ as a function of thickness and field-effect doping, complemented by electrical transport measurements. The phase transition persists down to the monolayer limit, with the critical temperature increasing as thickness decreases. In bilayer samples, both electron and hole doping suppress the insulating state, with electron doping lowering and hole doping raising the transition temperature. Importantly, the quasi-elastic scattering, previously attributed to excitonic fluctuations, evolves monotonically across the entire doping range, inconsistent with the expected suppression of excitonic correlations by Coulomb screening. These findings rule out a dominant excitonic mechanism and instead point to a coupled electronic and structural phase transition, whose stability is tunable by carrier doping. Our doping-based approach offers a general strategy for evaluating the role of excitonic effects in candidate excitonic insulators.
\end{abstract}

\maketitle

\noindent
In semiconductors or semimetals, unscreened Coulomb interaction can lead to the formation of bound electron-hole pairs known as excitons. If the exciton binding energy exceeds the band gap or band overlap, these composite bosons may condense at low temperature into a macroscopic quantum state---an excitonic insulator (EI)~\cite{Mott1961,Blatt1962,Keldysh1965,Cloizeaux1965,Jerome1967}. As a distinct type of correlated insulator, the EI has attracted sustained interest both as a platform to study electronically driven phase transitions and as a potential host for macroscopic quantum coherence, including superfluid-like behavior~\cite{Eisenstein2004,Wu2024}. While significant advances have been made in engineered bilayer systems with spatially separated electrons and holes~\cite{Eisenstein2014,Fogler2014,Du2017,Liu2017,Li2017,Wang2019,Ma2021,Rickhaus2021,Gu2022,Chen2022,Zhang2022}, realizing an EI phase in natural bulk crystals remains experimentally challenging. In addition to the scarcity of candidate materials, a major obstacle is the lack of unambiguous experimental signatures unique to exciton condensation~\cite{Cercellier2007,Kogar2017,Jia2022,Sun2022,Gao2023,Song2023,Gao2024,Zhang2024,Huang2024}.

A key complication in bulk EI candidates is the ubiquitous presence of electron-phonon coupling, which makes it difficult to isolate the electronic origin of the insulating ground state. For example, in the prototypical EI candidate 1$T$-TiSe$_2$, exciton condensation has been proposed to drive the emergence of a charge-density wave (CDW)~\cite{Cercellier2007,Kogar2017}. However, the CDW ordering wavevector coincides with a lattice distortion involving a soft phonon mode, making it difficult to disentangle electronic and lattice contributions~\cite{Wezel2010,Bianco2015}. In contrast, Ta$_2$NiSe$_5$ is a direct-gap semiconductor, avoiding complications from finite-wavevector CDWs. Several experimental works support an EI state in this material, including the flattening of the valence band top revealed by angle-resolved photoemission spectroscopy (ARPES)~\cite{Wakisaka2009,Seki2014}, a dome-shaped electronic phase diagram centered at zero band gap~\cite{Lu2017}, optical signature of exciton-phonon coupling~\cite{Larkin2017}, and ultrafast optical modulation of the insulating state~\cite{Mor2017,Okazaki2018,Bretscher2021,Katsumi2023}. Nevertheless, a structural phase transition also occurs in this material~\cite{Nakano2018,Subedi2020,Jog2022}, raising the possibility that the insulating state is of structural rather than electronic origin~\cite{Lu2021,Baldini2023,Chen2023,Chen2023b}. Some theoretical works suggested that exciton condensation~\cite{Kaneko2013} or even an excitonic instability~\cite{Mazza2020} can drive this structural transition, whereas others showed that the structural transition, rather than excitonic effect, is required to open the experimentally observed band gap~\cite{Windgatter2021,Baldini2023,Chen2023,Tang2020}. This entanglement of electronic and structural effects is not unique to Ta$_2$NiSe$_5$ and continues to pose a challenge for studying EIs.

\begin{figure*}[t]
\centering
\includegraphics[width=0.8\linewidth]{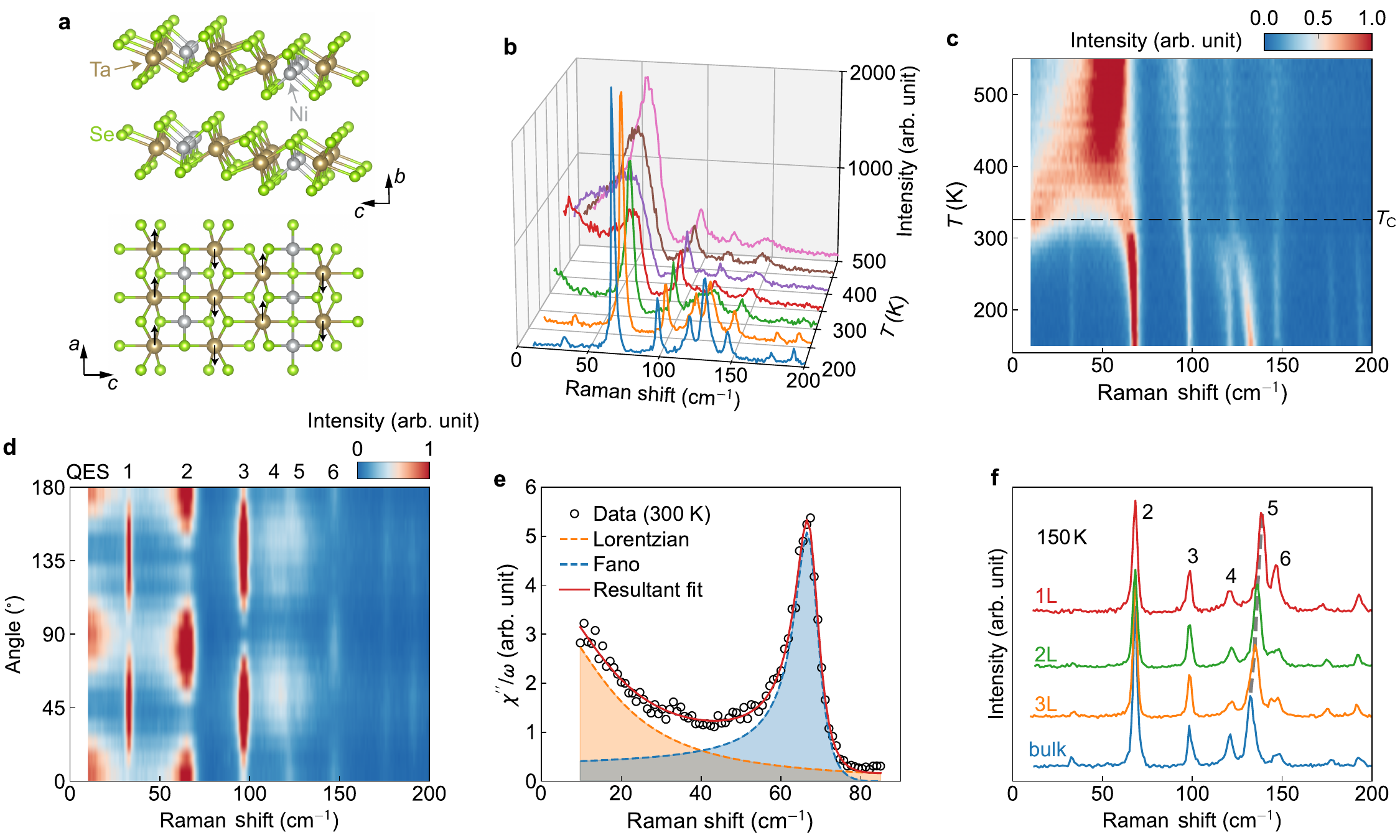}
\caption{\textbf{Crystal structure and Raman characteristics of Ta$_2$NiSe$_5$.} \textbf{a} Crystal structure of Ta$_2$NiSe$_5$ in the orthorhombic phase. The arrows in the lower part indicate atomic displacement in the monoclinic phase. \textbf{b}, \textbf{c} Temperature-dependent Raman spectra (\textbf{b}) and the corresponding intensity color plot (\textbf{c}) of bulk Ta$_2$NiSe$_5$. \textbf{d} Polarization-angle-dependent Raman intensity color plot of bulk Ta$_2$NiSe$_5$ at 300~K. \textbf{e} Fitting analysis of the low-energy Raman conductivity of bulk Ta$_2$NiSe$_5$ at 300~K. \textbf{f} Thickness-dependent Raman spectra of Ta$_2$NiSe$_5$ at 150~K. All data were collected in the crossed polarization ($ac$) configuration.}
\label{Fig1}
\end{figure*}

In this work, we propose an experimental approach to assess the excitonic nature of an insulating state by tracking its evolution under carrier doping. Since excitons are central to the EI state, and since carrier doping enhances Coulomb screening, possible hallmarks of an EI should be suppressed by both electron and hole doping. This anticipated doping dependence serves as a useful criterion for evaluating the role of excitonic effects. We implement this approach in atomically thin Ta$_2$NiSe$_5$, employing field-effect gating to achieve continuous and reversible doping. By monitoring the Raman response, we find that the quasi-elastic scattering (QES), previously attributed to excitonic fluctuations, evolves monotonically with doping across a wide range on both the electron and hole sides. This observation is inconsistent with expectations from an excitonic mechanism but supports a scenario in which the insulating state originates from a structural phase transition driven by electron-phonon coupling~\cite{Windgatter2021} and tunable via carrier density.

\vspace{2mm}
\noindent
\textbf{Results}\\
\noindent
\textbf{Coupled electronic and phononic excitations}\\
Bulk Ta$_2$NiSe$_5$ consists of monolayers stacked along the $b$-axis via van der Waals interactions (Fig.~1a), crystallizing in an orthorhombic structure (space group $Cmcm$) at high temperature~\cite{Sunshine1985,DiSalvo1986}. Within each monolayer, Ta and Ni atoms are coordinated by Se atoms in approximately octahedral and tetrahedral geometries, respectively, forming chain-like motifs along the $a$-axis. Below $T_{\mathrm{C}}=328$~K, the system undergoes a second-order phase transition into a monoclinic phase (space group $C2/c$), associated with a putative EI state. The two Ta chains flanking each Ni chain exhibit shear-like displacements in opposite directions (see arrows in Fig.~1a), breaking the in-plane mirror symmetries and giving rise to an antiferroelectric or ferro-rotational order~\cite{Nakano2018,Subedi2020,Jog2022}.

\begin{figure*}[t]
\centering
\includegraphics[width=\linewidth]{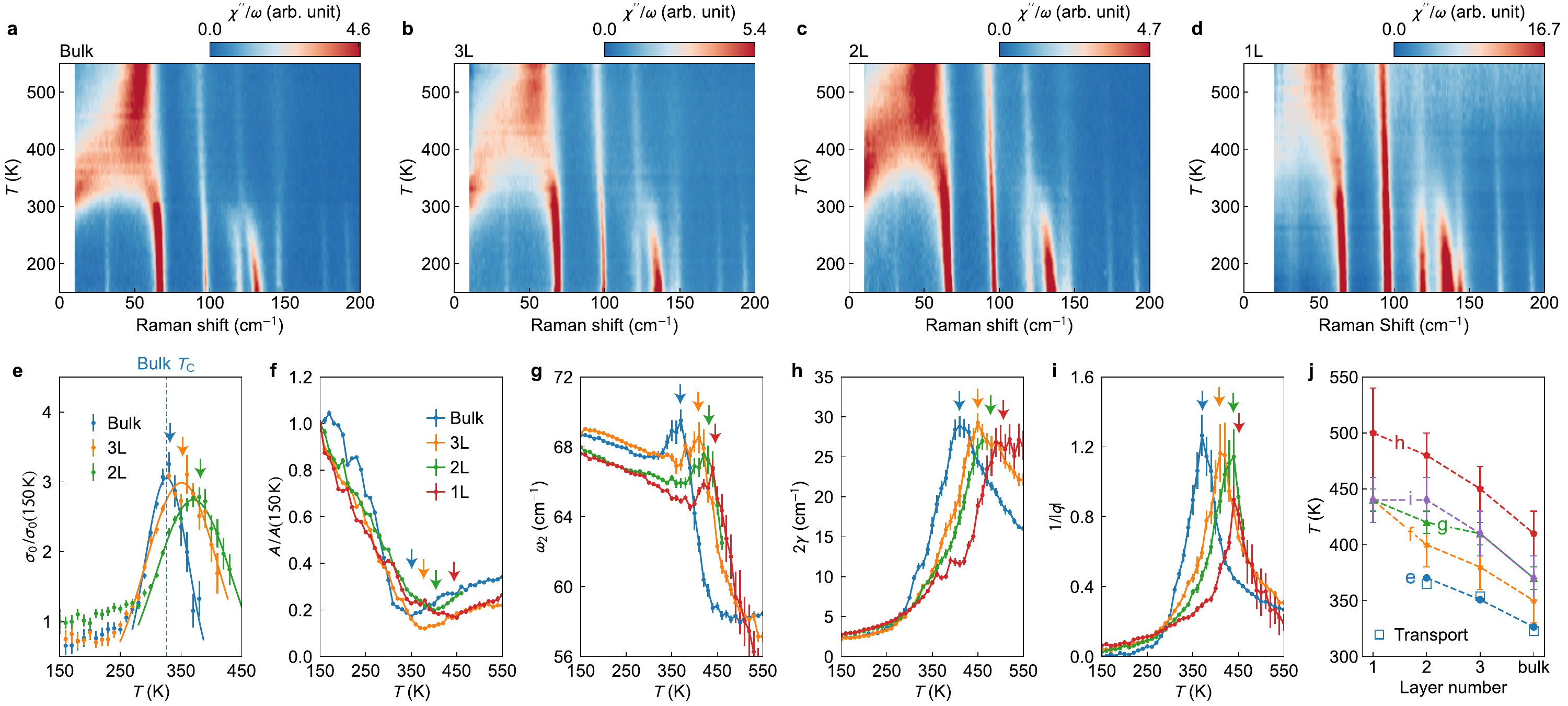}
\caption{\textbf{Thickness and temperature dependent Raman conductivity.} \textbf{a--d} Temperature-dependent Raman conductivity color plot for bulk, trilayer, bilayer, and monolayer samples. \textbf{e} Temperature dependence of the QES spectral weight. High-temperature data points are missing because the QES signal becomes too weak for reliable fitting. The solid lines are Gaussian fits. \textbf{f--i} Temperature dependence of the amplitude, frequency, full width, and $1/|q|$ of mode 2. \textbf{j} Thickness dependence of the temperature corresponding to the maximum or minimum values marked by arrows in \textbf{e--i} (filled symbols). The open squares are $T_{\mathrm{C}}$ values estimated from electrical transport measurements. Error bars in \textbf{e--i} are standard deviations obtained from fitting analysis, and those in \textbf{j} represent uncertainties in determining the characteristic temperatures. All data were collected in the crossed polarization ($ac$) configuration.}
\label{Fig2}
\end{figure*}

This phase transition is clearly manifested in the temperature-dependent Raman spectra (Fig.~1b--c), which show drastic changes both in the low-energy QES and in selected phonon modes. Near $T_{\mathrm{C}}$, the QES is strongly enhanced and hybridizes with the phonon mode at approximately 70~cm$^{-1}$, and the coupling of these two features is enabled by their compatible symmetry ($B_{\text{2g}}$ in the orthorhombic phase and $A_{\text{g}}$ in the monoclinic phase)~\cite{Kim2021,Volkov2021,Yan2019}, as further confirmed by polarization-angle dependent Raman measurements (Fig.~1d). Following Ref.~\cite{Kim2021}, we label the phonon modes numerically in order of increasing frequency. The crossed polarization ($ac$) configuration is adopted with the polarization angle that maximizes the intensity of the QES and mode 2, as they are particularly sensitive to the phase transition. Indeed, both mode 2 and mode 5 have been shown to exhibit shearing vibration of the Ta atoms along the chain direction, which matches the form of the monoclinic lattice distortion~\cite{Windgatter2021}.

QES was previously observed in iron pnictide superconductors~\cite{Gallais2013} and magnetic materials~\cite{Sandilands2015,Glamazda2016,Kim2019}, where it was attributed to charge nematic fluctuations and magnetic fluctuations, respectively. In Ta$_2$NiSe$_5$, however, since magnetic order is absent~\cite{DiSalvo1986}, no evidence of electronic nematicity was reported, and all expected Raman-active phonon modes can be accounted for~\cite{Kim2021}, the QES was attributed to an electronic origin~\cite{Kim2021,Volkov2021}. The QES exhibits a Lorentzian lineshape centered at zero frequency, resembling that observed in nematic and magnetic systems, which motivates a similar analysis framework~\cite{Gallais2013,Sandilands2015,Glamazda2016,Kim2019}. The measured Raman intensity $I(\omega)$ is converted to the Raman susceptibility, $\chi^{\prime\prime}(\omega) \propto I(\omega)/[n(\omega,T) + 1]$, where $n(\omega, T)$ is the Bose-Einstein distribution at temperature $T$. The Raman conductivity, defined as $\chi^{\prime\prime}(\omega)/\omega$, is modeled as a Lorentzian function centered at zero frequency, $\sigma_0 \Gamma / (\omega^2 + \Gamma^2)$, where $\sigma_0$ represents the spectral weight and $\Gamma$ is the half-width of the QES peak. The coupling between the QES and phonon mode 2 leads to a Fano lineshape for the latter, modeled by the expression $\frac{A}{1+q^2}\frac{(\epsilon+q)^2}{\epsilon^2+1}$, where $\epsilon = (\omega - \omega_2)/\gamma$, and $A$, $\omega_2$, and $\gamma$ denote the amplitude, frequency, and half-width of the phonon mode, respectively. The parameter $1/|q|$ quantifies the coupling strength between the phonon and the QES. A model incorporating both the Lorentzian and the Fano terms provides an excellent fit to the spectra at all temperatures where QES is present. A representative fit at 300 K is shown in Fig.~1e, with results at other temperatures provided in Supplementary Note~1. This analysis offers a robust method for quantitatively tracking the evolution of both the QES and mode 2 under varying experimental conditions.

\begin{figure*}[t]
\centering
\includegraphics[width=0.7\linewidth]{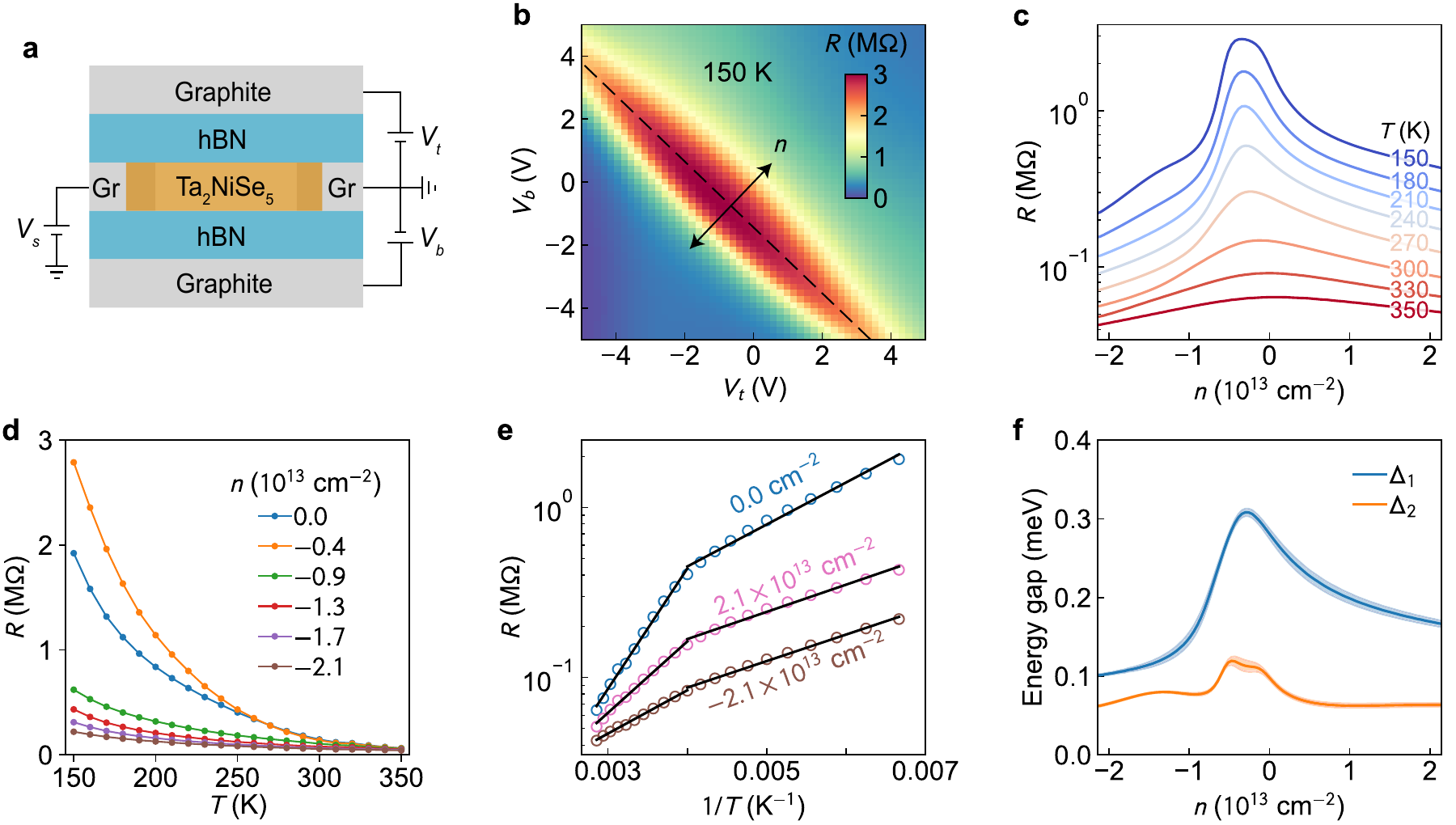}
\caption{\textbf{Doping dependent electrical transport properties of bilayer Ta$_2$NiSe$_5$.} \textbf{a} Schematic structure of the dual-gate bilayer Ta$_2$NiSe$_5$ device D17. Gr: graphite electrodes. \textbf{b} Dual-gate mapping of the resistance measured at 150~K. The dashed lines marks the charge neutrality and the arrows denote pure doping. \textbf{c} Doping-dependent resistance at selected temperatures. \textbf{d} Temperature-dependent resistance at charge neutrality and selected hole dopings. \textbf{e} Arrhenius plot of the resistance at typical doping levels. The solid lines are fits to the thermally activated temperature dependence over two temperature ranges. \textbf{f} Doping dependence of the activation gap. The shaded areas represent error bars obtained from the fitting analysis.}
\label{Fig3}
\end{figure*}

\vspace{2mm}
\noindent
\textbf{Thickness dependence}\\
Atomically thin Ta$_2$NiSe$_5$ samples were obtained by mechanical exfoliation. The number of layers was determined based on optical contrast and verified using atomic force microscopy (Supplementary Note~2). The frequency of mode 5 varies systematically with thickness (Fig.~1f), assisting the identification of the layer number. The persistence of sharp phonon peaks in samples down to the monolayer limit indicates good sample quality, in contrast to an earlier study where Raman signals were undetectable for samples thinner than four layers, likely due to degradation~\cite{Kim2016}.

We performed systematic temperature-dependent Raman measurements on samples with varying thickness. Figure~2a–d reveals that the phase transition persists down to the monolayer limit. Notably, $T_{\mathrm{C}}$ increases with decreasing thickness, as indicated by the temperature at which the QES reaches its maximum intensity and by the coexistence of modes 4 and 5 persisting to higher temperatures (see Supplementary Note~3 for mode analysis). Analysis of the QES and mode 2 using the method described above yields the parameters shown in Fig.~2e–i. For the bulk sample, the QES spectral weight $\sigma_0$ is maximized at the $T_{\mathrm{C}}$ determined from resistance measurements~\cite{Lu2017,DiSalvo1986}, strongly supporting its electronic origin. This maximum is retained in atomically thin samples and is used to extract a thickness-dependent $T_{\mathrm{C}}$ by fitting the QES temperature dependence above 250~K to a Gaussian function, which shows a systematic enhancement as the thickness decreases (blue circles in Fig.~2j). The monolayer sample exhibits enhanced background scattering when heated above 400~K, likely due to laser-induced degradation at elevated temperatures. Its spectral weight for the QES at high temperature and $T_{\mathrm{C}}$  cannot be reliably analyzed for this reason. The data shown in Fig.~2d have been corrected for this background contribution, as detailed in Supplementary Note 4. This issue was not observed in bilayer or thicker samples.

Fitting parameters associated with mode 2, including its amplitude $A$, frequency $\omega_2$, linewidth $2\gamma$, and Fano asymmetry $1/|q|$ are shown in Fig.~2f–i. The amplitude shows a minimum, whereas the other three parameters show a maximum near $T_{\mathrm{C}}$. These trends are consistent with previous bulk measurements~\cite{Kim2021}. The characteristic temperatures corresponding to the extrema in the parameters systematically increase as thickness is reduced, as marked by arrows in Fig.~2f–i and summarized in Fig.~2j, corroborating the enhancement of $T_{\mathrm{C}}$ in atomically thin samples. This finding contrasts with an earlier report of $T_{\mathrm{C}}$ reduction upon decreasing thickness~\cite{Kim2016}. Quantifying $T_{\mathrm{C}}$ from mode 2 parameters, however, is less reliable, as they show different characteristic temperatures even for the same sample. This could be due to the additional effect of lattice anharmonicity~\cite{Windgatter2021}, which becomes enhanced at high temperature and near structural phase transitions. Another source of discrepancy between the different criteria for determining $T_{\mathrm{C}}$ using mode 2 parameters arises from the data analysis itself. In principle, this could be remedied by employing a fully self-consistent analysis that couples the QES and the bare phonon~\cite{Volkov2021}. However, for Ta$_2$NiSe$_5$, such an approach has been shown to have only a minimal influence on the analyzed results~\cite{Kim2021}. We further performed resistance measurements on samples with varying thickness (Supplementary Note 5). The kink in the temperature derivative of the resistance as an indication of the phase transition~\cite{Kim2016} was observed in all measured samples, although it becomes weaker in atomically thin samples. The corresponding $T_{\mathrm{C}}$ values are plotted in Fig.~2j as open symbols, which confirm the increased $T_{\mathrm{C}}$ as the sample thickness decreases.

\begin{figure*}[t]
\centering
\includegraphics[width=0.8\linewidth]{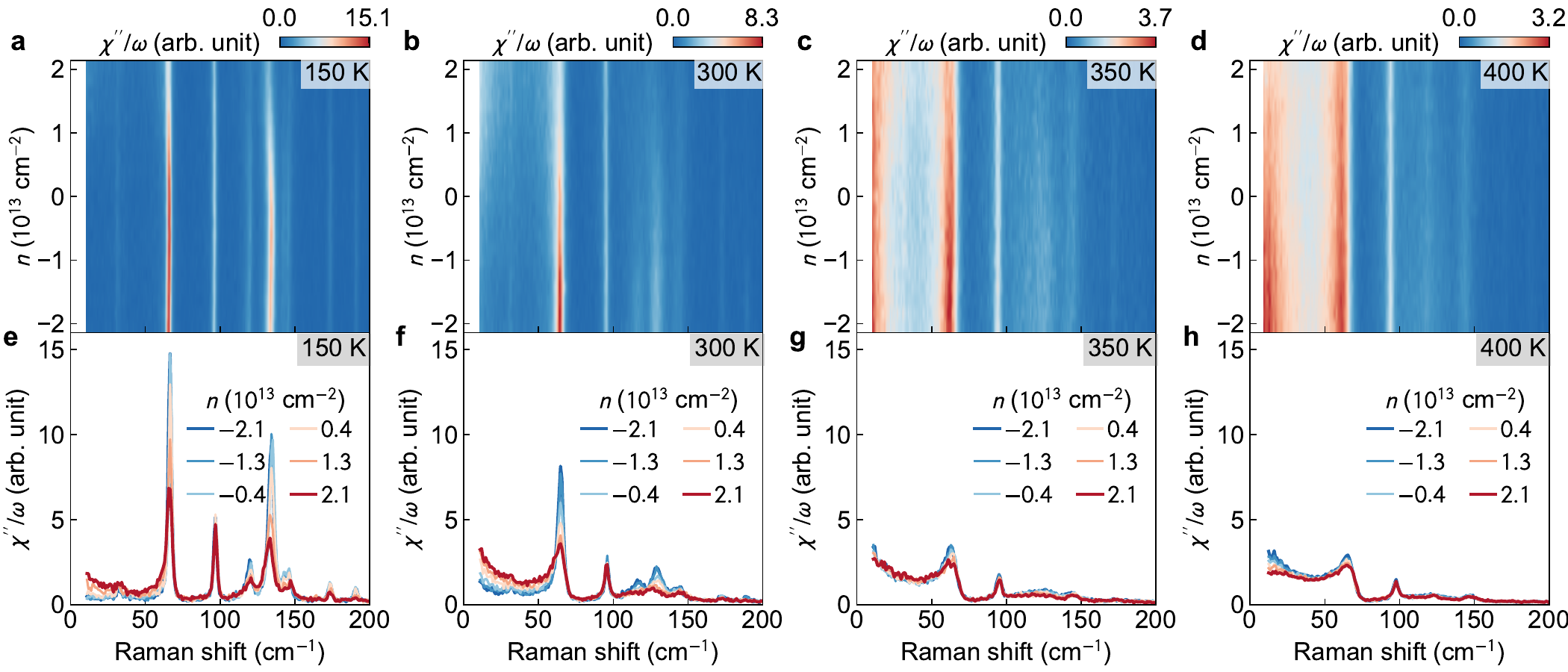}
\caption{\textbf{Doping dependence of the Raman conductivity in bilayer Ta$_2$NiSe$_5$.} \textbf{a--d} Doping-dependent Raman conductivity color plots for the bilayer sample D17 at selected temperatures. \textbf{e--h} Representative spectra corresponding to \textbf{a}--\textbf{d} at selected doping levels. All data were collected in the crossed polarization ($ac$) configuration.} 
\label{Fig4}
\end{figure*}

\vspace{2mm}
\noindent
\textbf{Effect of doping on electrical transport}\\
Having established the phase transition in atomically thin samples, we now investigate how it evolves under carrier doping. To this end, we integrated bilayer Ta$_2$NiSe$_5$ into a dual-gate field-effect transistor configuration, as illustrated schematically in Fig.~3a. A typical device image is shown in Supplementary Note 5. Two-probe resistance measurements were performed at 150 K while sweeping both the top-gate voltage ($V_t$) and bottom-gate voltage ($V_b$). As shown in Fig.~3b, the resistance peaks along a line corresponding to charge neutrality. This charge-neutrality line is slightly offset from the origin, possibly due to extrinsic doping. To focus on the effect of pure carrier doping, we follow the $V_t$-$V_b$ relation that ensures cancellation of the gate-induced electric field (see Methods), as marked by the arrowed line in Fig.~3b. Both electron and hole doping progressively suppress the insulating state, as seen in the resistance versus doping curve in Fig.~3c. Here, $n$ denotes the sheet carrier density, with positive (negative) values representing electron (hole) doping. The resistance exhibits a maximum at charge neutrality, which persists to high temperature but becomes thermally broadened. The maximum doping achieved amounts to approximately 0.06$e$/unit cell, estimated using the applied gate voltages and the thickness of the hexagonal boron nitride (hBN) dielectric layers.

\begin{figure*}[t]
\centering
\includegraphics[width=0.8\linewidth]{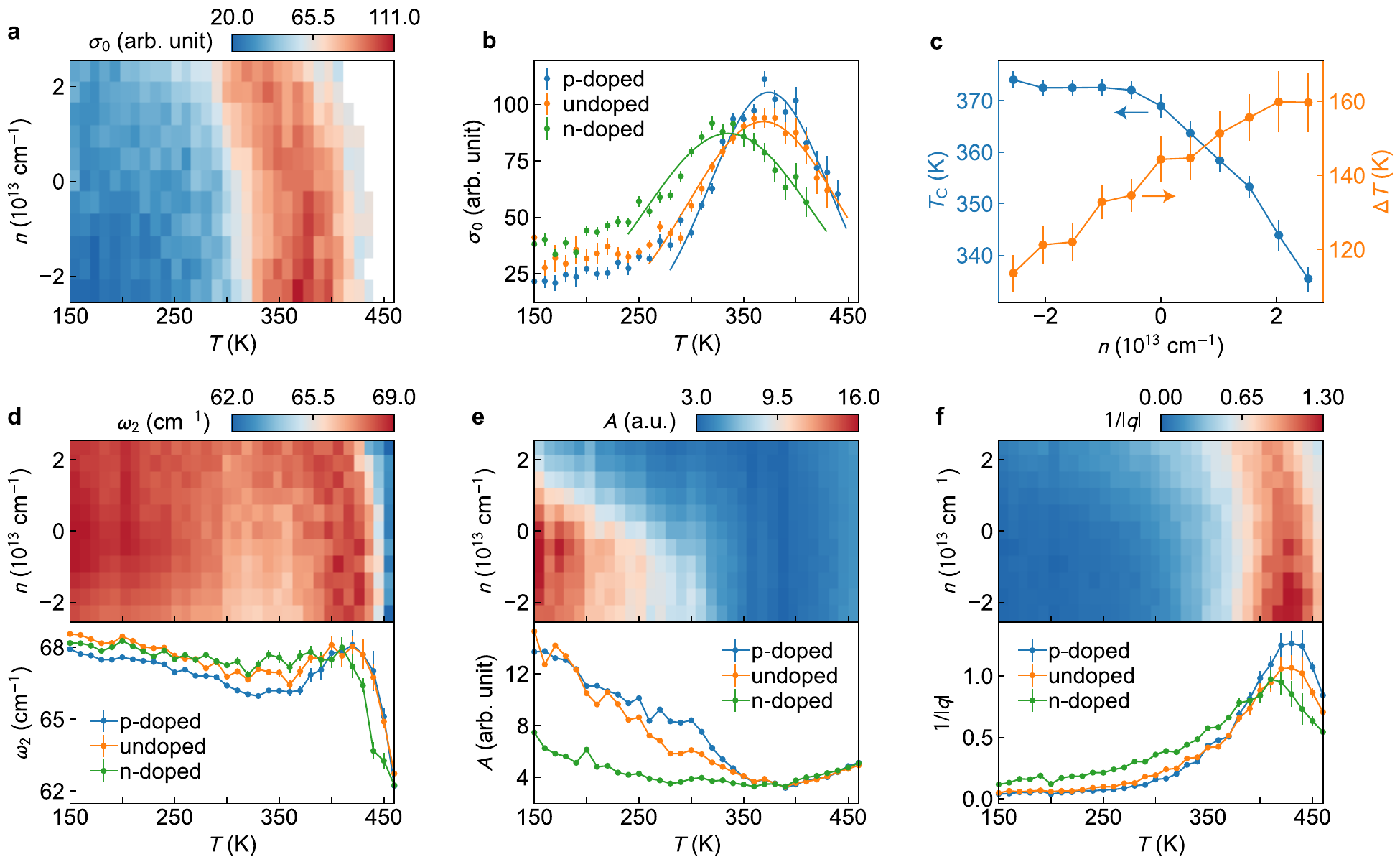}
\caption{\textbf{Doping and temperature dependent QES and mode 2 parameters for bilayer Ta$_2$NiSe$_5$ device D50.} \textbf{a}, Color map showing the doping and temperature dependence of the QES spectral weight. \textbf{b}, Line cuts along the temperature axis at zero doping (undoped), the maximum electron doping (n-doped), and the maximum hole-doping (p-doped) in \textbf{a}. The solid lines are Gaussian fits to the data above 250~K, with the fitted peak center corresponding to the $T_{\mathrm{C}}$ and the full width at half maximum characterizing the temperature spread ($\Delta T$) of the QES. \textbf{c} Doping dependence of $T_{\mathrm{C}}$ (left axis) and $\Delta T$ (right axis). \textbf{d}--\textbf{f}, The upper panels show the doping and temperature dependent color maps of the frequency, amplitude, and $1/|q|$ for mode~2. The lower panels are the corresponding line cuts at the same doping levels as those in \textbf{b}. Error bars in \textbf{b--f} are standard deviations obtained from fitting analysis.}
\label{Fig5}
\end{figure*}

The temperature dependence of resistance at selected hole doping densities is shown in Fig.~3d, revealing insulating behavior across the accessible doping range. The electron-doped side exhibits qualitatively similar behavior. An Arrhenius plot of the resistance (Fig.~3e) reveals two distinct thermal activation regimes. By fitting the temperature-dependent resistance to $R\propto \exp{(\Delta/2k_B T)}$, where $k_B$ is the Boltzmann constant, we extract two energy gaps: a larger gap, $\Delta_1$, which dominates above~250 K, and a smaller gap, $\Delta_2$, which governs transport at lower temperatures. Their doping dependence is summarized in Fig.~3f. While $\Delta_1$ is strongly doping-dependent and decreases with increasing carrier density, $\Delta_2$ remains largely unchanged above $10^{13}$~cm$^{-2}$ for both electron and hole doping. The smaller gap likely arises from defect-related in-gap states, consistent with prior scanning tunneling spectroscopy (STM) studies of exfoliated Ta$_2$NiSe$_5$~\cite{Kim2016}. The values of $\Delta_1=0.3$~eV and $\Delta_2=0.1$~eV at charge neutrality are comparable to those reported in bulk samples~\cite{Kim2016,Lu2017,Lee2019}.

At first glance, the suppression of insulating behavior by both electron and hole doping is consistent with the screening of an excitonic gap by doped carriers~\cite{Jia2022}. However, such ambipolar doping dependence is not unique to EIs. For instance, in Bernal-stacked bilayer graphene featuring a non-excitonic band gap opened by an electric field, the insulating state is similarly suppressed by both electron and hole doping~\cite{Liu2021}. In that case, the evolution of the activation gap is effectively associated with a shift in the chemical potential. In bilayer Ta$_2$NiSe$_5$, the persistence of $\Delta_1$ up to doping levels of $10^{13}$~cm$^{-2}$, along with the gradual slowing of its suppression at higher doping, suggests that the gap formation is unlikely to be dominated by excitonic effects~\cite{Windgatter2021}.

\vspace{2mm}
\noindent
\textbf{Effect of doping on Raman scattering}\\
To further elucidate the effect of carrier doping, we examine the corresponding evolution in the Raman spectra. Figures 4a–d show doping-dependent Raman conductivity maps at selected temperatures, with representative spectra displayed in Fig.~4e–h. At 150 K, the QES is nearly absent at zero doping but is induced by electron doping. At 300 K, QES is already present at charge neutrality and becomes clearly enhanced (suppressed) by electron (hole) doping. At 350 K, close to the $T_{\mathrm{C}}$ estimated earlier, the doping dependence of the QES weakens. At 400 K, the QES exhibits a doping dependence opposite to that observed at lower temperatures. The doping tunability of the QES affirms its electronic origin. Previous Raman studies attributed the QES at high temperatures to excitonic fluctuations~\cite{Kim2021,Volkov2021}, an interpretation supported by theoretical predictions~\cite{Sugimoto2018} and ARPES evidence for preformed excitons above $T_{\mathrm{C}}$~\cite{Fukutani2021}. However, our results challenge this view. If the QES indeed originated from excitonic fluctuations, both electron and hole doping would be expected to suppress it, as enhanced Coulomb screening weakens the excitonic bound states. Contrary to this expectation, we observe a monotonic doping dependence, inconsistent with the excitonic fluctuation scenario.

Recent x-ray scattering experiments revealed pronounced diffuse scattering along the inter-chain direction in bulk Ta$_2$NiSe$_5$, peaking at $T_{\mathrm{C}}$ and persisting over a broad temperature range, which suggests dynamic lattice fluctuations associated with an inter-chain shear mode that freeze into static distortion in the monoclinic phase~\cite{Chen2023}. These fluctuations enable hybridization between Ta-derived conduction and Ni-derived valence bands, otherwise symmetry-forbidden in the orthorhombic phase~\cite{Watson2020}. Resonant inelastic x-ray scattering further supports this picture by revealing substantial Ta-Ni orbital hybridization above $T_{\mathrm{C}}$~\cite{Lu2021}, which can manifest as a pseudogap in the orthorhombic phase, as indicated by ARPES~\cite{Chen2023} and infrared spectroscopy~\cite{Larkin2017} measurements. The QES intensity in our Raman data also peaks at $T_{\mathrm{C}}$, indicating a direct link to such fluctuations and justifying using the QES to quantify them, analogous to its application for quantifying charge nematic fluctuations in iron-pnictide superconductors~\cite{Gallais2013}. 

More systematic doping dependent Raman measurements were performed at a finer temperature step in another bilayer device, which allows us to extract the temperature dependence of the QES spectral weight at different doping levels. The results, shown in Fig.~\ref{Fig5}a, are consistent with the doping dependence at selected temperatures described above. By fitting the temperature dependence of the QES above 250~K to a Gaussian function (Fig.~\ref{Fig5}b), $T_{\mathrm{C}}$ is quantified as the center of the Gaussian peak, and its doping dependence is shown in Fig.~\ref{Fig5}c. We estimate an enhancement of 5~K and reduction of 34~K under the maximum hole and electron doping of $2.6\times 10^{13}$~cm$^{-2}$, respectively. The reduction of $T_{\mathrm{C}}$ under electron doping is consistent with previous potassium dosing studies and DFT calculations, which show that electron doping tends to stabilize the orthorhombic phase~\cite{Chen2023}. The doping dependence of the QES at selected temperatures, as shown in Fig.~\ref{Fig4}, can therefore be attributed to the doping-induced modification of $T_{\mathrm{C}}$, which shifts the peak in the temperature dependence of the QES spectral weight (Fig.~\ref{Fig5}a--b). The spread of the QES temperature dependence, quantified as the full width at half maximum of the Gaussian peak, grows as $T_{\mathrm{C}}$ decreases (Fig.~\ref{Fig5}c), indicating a more extended fluctuation regime under electron doping. 
 
Phonon mode 2 also exhibits clear doping dependence, especially below $T_{\mathrm{C}}$ (Fig.~\ref{Fig4}). At 150 K, where the QES is weak, mode 2 softens and broadens with doping (Supplementary Note 6), which could be due to doping-induced enhancement of either the electron-phonon coupling or the electronic density of states at the chemical potential. Above $\sim$250~K and below $T_{\mathrm{C}}$, when the QES becomes more prominent, the doping dependence of mode 2 parameters becomes monotonic (Fig.~\ref{Fig5}d--f). Electron doping reduces the mode amplitude (Fig.~\ref{Fig5}e) but increases its frequency (Fig.~\ref{Fig5}d), $1/|q|$ (Fig.~\ref{Fig5}f) and linewidth (Supplementary Note 6), resembling the effect of increasing temperature just below $T_{\mathrm{C}}$ (Fig.~2). A similar trend is observed for mode 5 (Supplementary Note~6), further corroborating electron-doping-enhanced lattice fluctuations when approaching $T_{\mathrm{C}}$ from below. Both modes are sensitive to the structural phase transition and dominated by Ta atomic vibration~\cite{Windgatter2021}. They couple efficiently with electronic states at the bottom of the conduction band formed by Ta-$5d$ orbitals~\cite{Kaneko2013}, explaining the more significant doping effect on the electron-doped side (see more discussions in Supplementary Note 6).

\vspace{2mm}
\noindent
\textbf{Discussion}\\
The phase transition in bulk Ta$_2$NiSe$_5$ has been attributed to a soft zone-center $B_{\mathrm{2g}}$ phonon, as predicted by DFT~\cite{Subedi2020}. Raman spectroscopy can in principle detect such a mode. While K. Kim et al. observed no soft-mode behavior~\cite{Kim2021}, M.-J. Kim et al. identified mode 2 as a soft mode based on measurements up to 800~K~\cite{Kim2020}. In our data, mode 2 shows slight softening and broadening between 550~K and 450~K (Fig.~2a, g, h), but this trend is interrupted by the enhancement of the QES. Moreover, mode 2 retains a Fano lineshape across the orthorhombic phase, even when the QES is not well resolved (Fig.~1 and Supplementary Note 1), indicating persistent electronic fluctuations beyond the scope of DFT. These observations support a phase transition driven by entangled electronic and structural instabilities~\cite{Windgatter2021}, which persist to higher temperature at reduced dimensionality. This enhanced $T_{\mathrm{C}}$ upon thickness reduction cannot be attributed to enhanced excitonic effects in reduced dimensions, because otherwise $T_{\mathrm{C}}$ should have been suppressed by both electron and hole doping, contrary to the monotonic doping dependence observed here. Instead, the enhanced $T_{\mathrm{C}}$ could be due to increased electron-phonon coupling that stabilizes monoclinic lattice distortion up to higher temperatures, which remains to be substantiated by further experimental and theoretical studies.

Previous doping studies on bulk Ta$_2$NiSe$_5$ using alkali-metal deposition were limited to electron doping and inevitably introduced surface electric fields~\cite{Fukutani2019,Chen2020,Chen2023}. In contrast, our dual-gate field-effect setup enables independent tuning of carrier density and electric field, providing a cleaner probe of intrinsic doping effects. We observed clear doping-induced changes in the Raman response that are inconsistent with excitonic condensation, while electric field effects were minimal (Supplementary Note 7). The field-effect doping approach in the bottom-gate geometry is also compatible with techniques such as ARPES and STM, offering a versatile route to probe excitonic correlations in other van der Waals EI candidates.

\vspace{7mm}
\noindent
\textbf{Methods}\\
\noindent
\textbf{Sample and device preparation}\\
Ta$_2$NiSe$_5$ single crystals were synthesized using the flux method. Freshly cleaved surfaces of the bulk crystals were used for Raman characterization. Thin flakes of Ta$_2$NiSe$_5$ were exfoliated from the bulk crystal onto polydimethylsiloxane (PDMS) substrates and identified based on optical reflection contrast, performed in an argon atmosphere within a glovebox. The exact thickness was determined through a combination of optical contrast analysis, atomic force microscopy (AFM), and Raman spectroscopy (see Supplementary Note~1). For Raman measurements, the flakes were encapsulated by hexagonal boron nitride (hBN) layers with thickness of 5--15~nm on both sides. Stacks were assembled by using a polycarbonate (PC) film on a PDMS stamp to sequentially pick up each material. Afterward, the stacks were released onto a Si/SiO$_2$ substrates at 200~$^{\circ}$C, and immersed in chloroform for 10~minutes and isopropanol for 5~minutes to remove the PC film.

Dual-gate field-effect transistors were fabricated for doping-dependent studies using the same exfoliation and transfer techniques. First, Ti/Au electrodes were patterned on Si/SiO$_2$ substrates via direct write laser lithography and thermal evaporation. Then, graphite electrodes (serving as contacts for Ta$_2$NiSe$_5$), hBN gate dielectric, and the bottom-gate graphite electrode were sequentially picked up and released onto the Ti/Au electrodes. The stack was cleaned using an AFM in contact mode, with a contact force of 5–20 nN. Finally, the top-gate graphite electrode, hBN gate dielectric, and Ta$_2$NiSe$_5$ flake were picked up and transferred onto the bottom-gate stack.

\vspace{2mm}
\noindent
\textbf{Characterizations}\\
Raman spectroscopy was conducted in the back-scattering geometry with 532~nm laser excitation. An incident power of 0.2~mW was used for all samples, which was confirmed to cause negligible heating effect. The scattered light passed through Bragg notch filters before being collected by a grating spectrograph and a liquid-nitrogen-cooled charge-coupled device. Measurements were performed in a cryostat within the 150--350~K range and in a heating stage between 300--550~K. To ensure consistency between datasets, spectra in the overlapping 300--350~K range were compared and calibrated (see Supplementary Note~8). Polarization-angle dependent measurements were carried out using two polarizers and a half-wave plate. Two-probe resistance was measured with a lock-in amplifier, applying an excitation current of 10~nA. Gate voltages were applied using source meters. Pure doping is achieved by applying gate voltages along the arrowed line shown in Fig.~3b. The doped sheet carrier density is calculated as $n = \epsilon_r\epsilon_0(V_t/d_t+V_b/d_b)/e$, in which the dielectric constant of hBN is $\epsilon_r=3$, $\epsilon_0$ is the vacuum permittivity, $e$ is the electron charge, $V_t$ and $V_b$ are the top and bottom gate voltages, respectively, and $d_t$ and $d_b$ are the thickness of the corresponding hBN dielectric layers. The electric field along the charge neutrality line shown in Fig.~3b is $E= (V_t/d_t -V_b/d_b)/2$.

\vspace{3mm}
\noindent
\textbf{Acknowledgements}\\
We thank Cheng Chen for helpful discussions. X.X. acknowledges support from the National Key Research and Development Program of China (Grant No.~2024YFA1409100), the Natural Science Foundation of Jiangsu Province (Grant Nos. BK20231529 and BK20233001), the Fundamental Research Funds for the Central Universities (Grant No. 0204-14380233), and the National Natural Science Foundation of China (Grant Nos. 12474170). Y.G. acknowledges support from the National Key Research and Development Program of China (Grant No.~2024YFA1408400). K.W. and T.T. acknowledge support from the JSPS KAKENHI (Grant Nos. 20H00354 and 23H02052) and World Premier International Research Center Initiative (WPI), MEXT, Japan.

\vspace{3mm}
\noindent
\textbf{Author contributions}\\
X.X. conceived the project. K.W. (NJU) performed the experiments. Y.L. and Y.G. grew the Ta$_2$NiSe$_5$ crystals. K.W. (NIMS) and T.T. grew the hBN crystals. K.W. (NJU) and X.X. analysed the experimental data and interpreted the results. X.X. wrote the paper, with comments from all authors.

\vspace{3mm}
\noindent
\textbf{Corresponding authors}\\
Correspondences should be sent to Yanfeng Guo (\href{mailto:guoyf@shanghaitech.edu.cn}{guoyf@shanghaitech.edu.cn}) or Xiaoxiang Xi (\href{mailto:xxi@nju.edu.cn}{xxi@nju.edu.cn}).  

\vspace{3mm}
\noindent
\textbf{Competing interests}\\
The authors declare no competing interests.
\newpage

\end{document}